\documentclass[journal]{IEEEtran}

\usepackage{citesort}
\usepackage{subfigure}
\usepackage{graphicx,epstopdf}
\usepackage{epsfig}	
\usepackage{cite,graphicx,amsmath,amssymb}
\usepackage{comment}
\usepackage{amsthm}
\usepackage{lipsum}
\usepackage{color}
\usepackage{placeins}
\usepackage{float}
\usepackage{tabularx,colortbl}
\usepackage{url}                    
\usepackage{lscape}                 
\usepackage{subfigure}
\usepackage{mathrsfs}
\usepackage{caption2}
\usepackage{epstopdf}
\usepackage{framed}
\usepackage{xcolor}

\usepackage{multirow}
\usepackage{algorithm}
\usepackage{algorithmic}
\usepackage{flafter}

\newtheorem{theorem}{Theorem}

\makeatletter
\def\ScaleIfNeeded{%
\ifdim\Gin@nat@width>\linewidth \linewidth \else \Gin@nat@width
\fi } \makeatother

\begin{document}
\title{Throughput Analysis of Wireless Powered Cognitive Radio Networks with Compressive Sensing and Matrix Completion}

\author{Zhijin Qin,
        Yuanwei Liu,
        Yue Gao,
        Maged Elkashlan,
        and Arumugam Nallanathan

\thanks{Z. Qin, Y. Liu, Y. Gao and M. Elkashlan are with Queen Mary University of London, London, United Kingdom. (email:\{z.qin; yuanwei.liu; yue.gao; maged.elkashlan\}@qmul.ac.uk)}
\thanks{A. Nallanathan is with King's College London, London, United Kingdom. (email: arumugam.nallanathan@kcl.ac.uk)}
}

\maketitle

\begin{abstract}
In this paper, we consider a cognitive radio network in which energy constrained secondary users (SUs) can harvest energy from the randomly deployed power beacons (PBs). A new frame structure is proposed for the considered network. A wireless power transfer (WPT) model and a compressive spectrum sensing model are introduced. In the WPT model, a new WPT scheme is proposed, and the closed-form expressions for the power outage probability are derived. In compressive spectrum sensing model, two scenarios are considered: 1) Single SU, and 2) Multiple SUs. In the single SU scenario, in order to reduce the energy consumption at the SU, compressive sensing technique which enables sub-Nyquist sampling is utilized. In the multiple SUs scenario, cooperative spectrum sensing (CSS) is performed with adopting low-rank matrix completion technique to obtain the complete matrix at the fusion center. Throughput optimizations of the secondary network are formulated into two linear constrained problems, which aim to maximize the throughput of single SU and the CSS networks, respectively. Three methods are provided to obtain the maximal throughput of secondary network by optimizing the time slots allocation and the transmit power. Simulation results show that: 1) Multiple SUs scenario can achieve lower power outage probability than single SU scenario; and 2) The optimal throughput can be improved by implementing compressive spectrum sensing in the proposed frame structure design.\\
\\
{\bf Keywords: }Compressive sensing, low-rank matrix completion, spectrum sensing, sub-Nyquist sampling, wireless power transfer.

\end{abstract}

\section{Introduction}
\IEEEPARstart{E}{nergy} efficiency and spectrum efficiency are two critical issues in designing wireless networks. Recent developments in energy harvesting provides a promising technique to improve the energy efficiency in wireless networks. Different from harvesting energy from traditional energy sources (e.g., solar, wind, water, and other physical phenomena)~\cite{raghunathan2006emerging}, the emerging wireless power transfer (WPT) further underpins the trend of green communications by harvesting energy from radio frequency (RF) signals~\cite{le2008efficient}. Inspiring by the great convenience offering by WPT, several works have been studied to investigate the performance of different kinds of energy constraint networks~\cite{zhang2013mimo,huang2014enabling,ding2014power,yuanwei_JSAC_2015}. Two practical receiver architectures, namely a time switching receiver and a power splitting receiver, were proposed in a multi-input and multi-output (MIMO) system in~\cite{zhang2013mimo}, which laid a foundation in the recent research of WPT. In~\cite{huang2014enabling}, a new hybrid network architecture is designed to enable charging mobiles wirelessly in cellular networks. For cooperative systems, new power allocation strategies are proposed in a cooperative networks where multiple sources and destinations are communicated by an energy harvesting relay~\cite{ding2014power}. For non-orthogonal multiple access (NOMA) networks, in~\cite{yuanwei_JSAC_2015}, a new cooperative simultaneously wireless information and power transfer NOMA protocol is proposed with considering the scenario where all users are randomly deployed.

Along with improving energy efficiency through energy harvesting, cognitive radio (CR) technique can improve the spectrum efficiency and capacity of wireless networks through dynamic spectrum access~\cite{Xiao:2014}. Therefore, in order to design networks which are both spectrum and energy efficient, the secondary users (SUs) in CR networks can be equipped with the energy harvesting capability. For the SUs powered by energy harvested from wireless radio frequency, high sampling rate is difficult to be achieved. To overcome this issue, compressive sensing (CS), which was initially proposed in~\cite{Candes:2006}, is introduced to wideband spectrum sensing in~\cite{tian:2007} to reduce the power consumption at SUs. As the spectrum of interest is normally underutilized in reality~\cite{Force2002,Ofcom:2009}, it exhibits a sparse property in frequency domain, which makes sub-Nyquist sampling possible by implementing the CS technique at SUs. In addition, when dealing with matrices containing limited available entries, low-rank matrix completion (MC)~\cite{candes2009exact} was proposed to recover the complete matrix. As the sparse property of received signals can be transformed into low-rank property of the matrix constructed in the cooperative networks, in~\cite{Jia:2011}, the authors  proposed to apply joint sparsity recovery and low-rank MC to reduce the sensing and transmission requirements and improve the sensing performance in CR networks. In order to improve robustness against channel noise, a denoised algorithm was proposed in~\cite{Zhijin_TSP:2015} for compressive spectrum sensing at single SU and low-rank MC based spectrum sensing at multiple SUs.

\subsection{Related works}
Some throughput optimization works have been recently developed in wireless powered communication networks. The authors in~\cite{LingjieDuan:2015} considered the throughput maximization problem for both battery-free and battery-deployed cases by optimizing the time slots for energy harvesting and data transmission. In addition, recent work~\cite{park2013cognitive,Park:2013,Chung:2014} on the CR networks powered by energy harvesting mainly focuses on the spatial throughput optimization under various constraints. The authors in~\cite{park2013cognitive} considered CR networks with an energy-harvesting SU with infinite battery capacity. The goal is to determine an optimal spectrum sensing policy that maximizes the expected total throughput subject to an energy causality constraint and a collision constraint. In order to improve both energy efficiency and spectral efficiency, the authors in~\cite{Park:2013} considered a similar network model and the stochastic optimization problem is formulated into a constrained partially observable Markov decision process. At the beginning of each time slot, a SU needs to determine whether to remain idle so as to conserve energy, or to execute spectrum sensing to acquire knowledge of the current spectrum occupancy state. The throughput is maximized by the design of a spectrum sensing policy and a detection threshold. The authors in~\cite{Chung:2014} consider an energy constraint RF-powered CR network by optimizing the pair of the sensing duration and the sensing threshold to maximize the average throughput of the secondary network.

\subsection{Motivations and contributions}
The aforementioned works have played a vital role and laid solid foundation for fostering new strategies for frame structure design. However, in~\cite{LingjieDuan:2015}, spectrum sensing is not considered in the frame structure design. In addition, in~\cite{park2013cognitive,Park:2013,Chung:2014}, the proposed frame structure designs mainly aim to maximize the throughput by optimizing the threshold and time slots.  When considering the energy efficiency and spectrum efficiency, it is meaningful to introduce sub-Nyquist sampling to reduce the energy consumption at SUs in wireless powered CR networks. In this paper, we propose a new frame structure design for wireless powered CR networks with implementing sub-Nyquist sampling at SUs. The CS and MC techniques are adopted to perform the signal recovery at a remote fusion center (FC) for making decision on spectrum occupancy.  To the best of our knowledge, this is the first paper on the frame structure design employing the sub-Nyquist sampling rates based spectrum sensing in wireless powered CR networks.

The summarized contributions of this paper are illustrated as follows:
\begin{itemize}
  \item We propose a new frame structure for wireless powered CR networks which includes four time slots: energy harvesting, spectrum sensing, energy harvesting and data transmission. We introduce a WPT model and a spectrum sensing model with sub-Nyquist sampling for the considered networks.
  \item In the WPT model, we propose a new bounded WPT scheme where each SU selects a PB nearby with the strongest channel to harvest energy. We derive closed-form expressions for the power outage probability.
  \item In the spectrum sensing model, sub-Nyquist sampling is performed at each SU to reduce the energy consumption during spectrum sensing period. We consider two scenarios: single SU scenario and multiple SUs scenario. In the single SU scenario, we adopt CS recovery algorithm to obtain the original signal. In the multiple SUs scenario, we utilize MC algorithm to obtain complete matrix at the FC to perform cooperative spectrum sensing (CSS). When the signal recovery process is performed at the remote FC, energy harvesting can be performed again at the SUs locally in the third time slot.
  \item Throughput optimizations of the proposed frame structure are formulated into two linear constrained problems with the purpose of maximizing the throughput of a single SU and the whole cooperative networks, respectively. The formulated problems are solved by using three different methods to obtain the maximal achievable throughput respectively.
  \item  Simulation results show that the proposed frame structure design outperforms the traditional one in terms of throughput. It is noted that the multiple SUs scenario can achieve better outage performance than the single SU scenario.
\end{itemize}

\subsection{Organizations}
The rest of this paper is organized as follows. Section~II describes the considered WPT model and spectrum sensing model based on the proposed frame structure. Section III presents the throughput analysis of single SU scenario with applying CS technique. Section IV provides the throughput analysis of multiple SUs with adopting MC technique. Section V shows the numerical analyses of the considered network model with the optimized throughput of single SU and multiple SUs, respectively. Section VI concludes this paper.

\section{Network Model}\label{Network Model}
\subsection{Network description}
\begin{figure*}[t!]
    \begin{center}
        \includegraphics[width=5.1in]{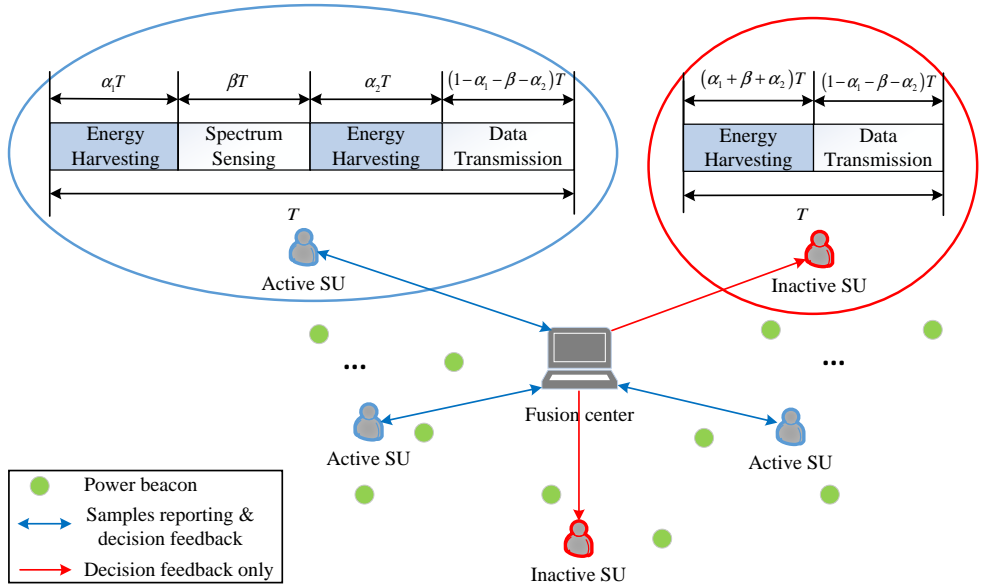}
        \caption{Proposed frame structure design with energy harvesting, spectrum sensing and data transmission.}
        \label{system_model}
    \end{center}
\end{figure*}

We consider a CR network, where SUs are energy constrained. The whole spectrum of interest can be divided into $I$ channels. A channel is either occupied by a primary user (PU) or unoccupied. Meanwhile, there is no overlap between different channels. The number of occupied channels is assumed to be $K$, where $K\le I$. Each SU is supposed to perform sensing on the whole spectrum. It is assumed that all SUs keep quiet as forced by protocols, e.g., at the media access control layer during spectrum sensing period. Thus the received signals only contain the signals of active PUs and channel noise. As shown in Fig.~\ref{system_model}, for each SU, it is assumed that the sensing and transmission can only be scheduled by utilizing energy harvested from power beacons (PBs). The spatial topology of all PBs are modeled using homogeneous poisson point process (PPP) $\Phi_p$ with density $\lambda_{p}$. Without loss of generality, we consider that a typical SU is located at the origin in a two-dimensional plane. Each SU is equipped with a single antenna and has a corresponding receiver with fixed distance. Each PB is furnished with $M$ antennas and maximal ratio transmission (MRT) is employed at PBs to perform WPT to the energy constrained SU. The energy harvesting channels are assumed to be quasi-static fading channels where the channel coefficients are constant for each transmission block but vary independently between different blocks. The spectrum of interest is wideband， and each SU performs wideband spectrum sensing to discover spectrum holes for data transmission. Once the spectrum holes are identified, SUs can start data transmission. It is assumed that the time of each frame is $T$. In the considered networks, single SU and multiple SUs scenarios are analyzed to achieve different throughput targets.

\begin{enumerate}
  \item Single SU scenario: in the considered spectrum sensing network with single SU, a frame period at a single SU includes four time slots as outlined in the blue oval in Fig.~\ref{system_model}: 1) energy harvesting time slot, in which each SU harvests the energy from PBs during the ${\alpha _1}T$ period, with $\alpha_1$ being the fraction of energy harvesting in one frame period; 2) spectrum sensing time slot, in which each SU performs sub-Nyquist sampling by applying CS techniques. The compressed measurements are then sent to a remote powerful FC during the $\beta T$ period by using the harvested energy during the ${\alpha _1}T$ period; 3) energy harvesting time slot for data transmission, in which each SU harvests the energy from PBs during the ${\alpha _2}T$ period, with ${\alpha _2}$ being the fraction of energy harvesting in one frame period. As the spectrum is typically underutilized in practice, signals received at each SU exhibits a sparse property. In this time slot, the original signals can be recovered based on the collected measurements. The sensing decisions are sent back to the corresponding SU at the end of the third time slot; and 4) data transmission slot, in which each SU performs data transmission during the $\left( {1 - {\alpha _1} - \beta  - {\alpha _2}} \right) T$ period.

  \item Multiple SUs scenario: as shown in Fig.~\ref{system_model}, in the considered networks with multiple SUs, named as cooperative spectrum sensing, SUs are spatially randomly distributed. The total number of participating SUs is $J \left( {J \ge 1} \right)$ in the CSS networks. Before performing spectrum sensing, each participating SU compare the harvested energy $E_{H_1}$ with the energy consumption for spectrum sensing $E_{S}=P_{H_1}\beta T$ at the first time slot. If $E_{H_1}$ is greater than $E_{S}$, the SU would continue performing spectrum sensing. This kind of SUs are named as active SUs. The frame structure of active SUs is same as that of single SU as described above. If $E_{H_1}$ is less than $E_S$, the SU would switch to energy harvesting model again and wait for the decision on spectrum occupancies from the FC before starting data transmission. Therefore, the framework structure of inactive SUs would only includes two time slots: $\left( {{\alpha _1} + \beta  + {\alpha _2}} \right)T$ for energy harvesting and the rest for data transmission. In the case of CSS, only measurements from active SUs are collected at the FC. The signals received at SUs exhibit a sparsity property that yields a low-rank matrix of compressed measurements at the FC. Therefore, the full information of spectrum occupancies can be obtained by adopting low-rank MC methods.
\end{enumerate}

\subsection{Wireless power transfer model}\label{Power transfer model}
We consider a bounded power transfer model with a protection zone with a radius $d_0$, which means that no PB is allowed to exist in this zone. If PBs are really close to the SU, the harvested energy would mathematically go to infinity~\cite{Venkataraman：2006}. It is assumed that the SU is battery-free~\cite{Liu2015TWC,yuanwei_JSAC_2015}, which means that there is no battery storage energy for future use  and all the harvested energy during energy harvesting time slots is used to perform spectrum sensing and data transmission in the current frame period.

We adopt a power transfer scheme where the SU selects the PB with the strongest channel to harvest energy. At each SU, the energy harvested from the selected PB in the first and the third time slots can be obtained as follows:
\begin{align}\label{EH Hk}
{E_{{H}}} = \mathop {\max }\limits_{p \in {\Phi _p},\left\| {{d_p}} \right\| \ge {d_0}} \left\{ {{{\left\| {{{\bf{h}}_p}} \right\|}^2}L\left( {{d_p}} \right)} \right\}\eta {P_p}\gamma T,
\end{align}
where $\gamma$ is the ratio of the time used for energy harvesting to the total time of a frame, $\eta$ is the power conversion efficiency at the SU, ${P_{p}}$ is the transmit power of PBs. Here, ${\bf{h}}_p$ is a ${\mathcal{C}^{M \times 1}}$ vector, whose entries are independent complex Gaussian distributed with zero mean and unit variance employed to capture the effects of small-scale fading between PBs and the SU. $L\left( {{d_p}} \right) = Ad_p^{ - \xi}$ is the power-law path-loss exponent. The path-loss function depends on the distance $d_p$, a frequency dependent constant $A$, and an environment/terrain dependent path-loss exponent $\xi \geq 2$. All the channel gains are assumed to be independent and identically distributed (i.i.d.).

For the active SUs, based on~\eqref{EH Hk}, the maximum sensing power at the SU is given by
\begin{align}\label{Power for sensing}
{P_{{H_1}}} = \frac{{{E_{{H_1}}}}}{{\beta T}} = \mathop {\max }\limits_{p \in {\Phi _p},\left\| {{d_p}} \right\| \ge d_0} \left\{ {{{\left\| {{{\bf{h}}_p}} \right\|}^2}L\left( {{d_p}} \right)} \right\}\frac{{\eta {P_p}{\alpha _1}}}{\beta }.
\end{align}

As energy can only be stored in the current frame, the total energy for data transmission is the sum of the remaining energy in the second slot and the energy harvested in the third time slot, which is given by
\begin{align}\label{Energy for transmission}
{E_{{T_2}}} =\left( {{E_{{H_1}}} - E_S} \right)+ {E_{{H_2}}},
\end{align}
where $P_s$ is the sensing power consumed at an SU.

Based on \eqref{EH Hk} and \eqref{Energy for transmission}, the corresponding power for data transmission is given by
\begin{align}\label{power for transmission}
{P_{{T_2}}}= \frac{{\mathop {\max }\limits_{{p \in {\Phi _p},\left\| {{d_p}} \right\| \ge d_0}} \left\{ {{{\left\| {{{\bf{h}}_p}} \right\|}^2}L\left( {{d_p}} \right)} \right\}\eta {P_p}\left( {{\alpha _1} + {\alpha _2}} \right) - {P_s}\beta }}{{1 - {\alpha _1} - \beta  - {\alpha _2}}}.
\end{align}

For those inactive SUs, the harvested energy before data transmission is given by
\begin{align}\label{EH3}
{E_{{H_3}}} = \mathop {\max }\limits_{p \in {\Phi _p},\left\| {{d_p}} \right\| \ge {d_0}} \left\{ {{{\left\| {{{\bf{h}}_p}} \right\|}^2}L\left( {{d_p}} \right)} \right\}\eta {P_p}{\left( {{\alpha _1} + \beta  + {\alpha _2}} \right)} T.
\end{align}

Based on~\eqref{EH3}, the maximum sensing power at the SU can be expressed as
\begin{align}\label{PH3}
{P_{{H_3}}} = \frac{{{E_{{H_3}}}}}{{\left( {1 - {\alpha _1} - \beta  - {\alpha _2}} \right)T}}.
\end{align}

\subsection{Spectrum sensing model with sub-Nyquist sampling}\label{CSS model}
In the considered spectrum sensing model, when $J=1$, it becomes a single node scenario. At the $j$th SU ($SU_j$) in the considered network, the received signals can be expressed as:
\begin{align}
r_{j}\left( t \right) = h_{j}\left( t \right)  *  s\left( t \right) + n_{j}\left( t \right),
\end{align}
where $s\left( t \right) \in {C^{n \times 1}}$ refers to the transmitted primary signals in time domain, and $h_{j}\left( t \right)$ is the channel gain between the transmitter and receiver, and $n_{j}\left( t \right) \sim \mathcal{CN}(0,{\sigma ^2}{\textbf{I}_n})$ refers to Additive White Gaussian Noise (AWGN) with zero mean and variance ${\sigma ^2}$.

Based on the Nyquist sampling theory, the sampling rate is required to be at least twice of the bandwidth. It leads to high energy cost which would be challenging to the energy constrained SUs in a CR network. It is noticed that the transmitted signal $s\left( t \right)$ exhibits a sparse property in frequency domain as a large percentage of spectrum is normally underutilized in practice. This sparse property makes the sub-Nyquist sampling at SUs by adopting CS techniques. After CS technique is implemented at SUs, the compressed measurements collected at ${SU}_{j}$ can be expressed as
\begin{align}
 x_j ={\Phi _{j} }r_{j}\left( t \right)= {\Phi _{j} }{\mathcal{F}_{j}^{ - 1}}{r_{j}\left( f \right)}={\Theta _j}\left( {{s}\left( f \right) + {n_j}\left( f \right)} \right),
\label{received}
\end{align}
where $\Phi _j  \in {C^{\Lambda \times n}}$ $(\Lambda< n)$ is the measurement matrix utilized to collect the compressed measurements $x_j \in {C^{\Lambda \times 1}}$, and ${{s}\left( f \right)}$ and ${{n_j}\left( f \right)}$ refer to the transmitted signals and the AWGN received at in frequency domain. The compression ratio at SUs is defined as $\kappa  = \frac{\Lambda }{n},\left( 0 \le \kappa  \le 1 \right)$. In addition, $\Theta_j  = {\Phi_j}\mathcal{F}_{j}{^{ - 1}}$, where $\mathcal{F}_{j}{^{ - 1}}$ is inverse discrete Fourier transform (IDFT) matrix.

After the compressed measurements are collected, SUs will send them to a remote FC by a error-free reporting channel. In both the single node and multiple nodes scenarios, FC is proposed to perform signal recovery efficiently. By adopting such a powerful FC, energy constrained SUs can get rid of signal recovery process and continue harvesting energy from PSs. At the FC, the compressed measurements $X$ can be expressed as
\begin{align}
X = \Theta {R_f}=\Theta \left({S_f}+{N_f}\right),
\label{collected_measurements_at_FC}
\end{align}
where ${R_f} = \left[ {{r_1}\left( f \right),{r_2}\left( f \right), \ldots ,{r_J}\left( f \right)} \right]$ which is in size of $n \times J$, and $S_f$ and $N_f$ refer to the matrix constructed by transmitted signals and AWGN. In addition, the measurement matrix is a diagonal matrix $\Theta  = diag\left( {{\Theta _1},{\Theta _2}, \ldots ,{\Theta _J}} \right)$ in size of $\Omega \times N$ with $\Omega =\Lambda \times J$ and $N=n\times J$. After the compressed measurements are collected at the FC, signal recovery can be formulated as a convex optimization problem. As aforementioned, the considered network becomes a single node case when $J=1$. Then the existing algorithm for CS can be utilized to recover the original signals. In the cooperative networks ($J>1$), existing algorithms for low-rank MC can be implemented to complete the matrix. It has been proved that the exact signal or matrix can be recovered if the number of available measurements are no less than the minimum bound. With enough number of measurements, exact transmitted signals can be obtained at the FC. Then energy detection is adopted to determine the spectrum occupancies, in which the decision is made by comparing the energy of recovered signal with a threshold defined as~\cite{Ying-chang:2008}
\begin{align}\label{threshold}
\lambda  = \left( {\sigma _s^2 + {\sigma ^2}} \right)\left( {1 + \frac{{{Q^{ - 1}}\left( {{{\bar P}_d}} \right)}}{{\sqrt {n/2} }}} \right).
\end{align}
Once the sensing decisions are determined at the FC, they would be sent back to the participating SUs in the CSS network by the reporting channel to start the data transmission period.

\section{Throughput Analysis of a Single Secondary User}\label{Throughput analysis_section}
In this section, the closed form expressions are derived for the power outage probability of spectrum sensing and data transmission, respectively. With CS technique implemented, the analysis with the target of maximizing throughput of each SU locally is provided in this section.

\subsection{Power outage probability analysis}\label{Power Outage Probability}
We assume there exists a threshold transmit power, below which the spectrum sensing in the second slot or the data transmission in the fourth slot cannot be scheduled. We introduce power outage probability, i.e., probability that the harvested energy is not sufficient to perform spectrum sensing or carry out the data transmission at a certain desired quality-of-service (QoS) level. In practical scenarios, we expect a constant power for the data transmission. Therefore, we also denote the power threshold $P_s$ as the sensing power in the second slot and $P_t$ as the transmit power of the SU, respectively. The following theorem provides the exact analysis for the power outage probability at the single SU scenario.
\begin{theorem}
The power outage probability of spectrum sensing $P_s ^{out}$ in the second time slot and the power outage probability of data transmission $P_t ^{out}$ in the fourth time slot can be expressed in closed-form as follows:
\begin{align}\label{optimal power outage}
P_\zeta ^{out} = \exp \left[ { - \frac{{\pi {\lambda _p}\delta }}{{\mu _\zeta ^\delta }}\sum\limits_{m = 0}^{M - 1} {\frac{{\Gamma \left( {m + \delta ,{\mu _\zeta }d_0^\xi } \right)}}{{m!}}} } \right],
\end{align}
where $\zeta  \in \left( {s,t} \right)$, ${\mu _s} = \frac{{\beta {P_s}}}{{\eta {P_p}A{\alpha _1}}}$, ${\mu _t} = \frac{{{P_t}\left( {1 - {\alpha _1} - \beta  - {\alpha _2}} \right) + {P_s}{\beta}}}{{\eta {P_p}A\left( {{\alpha _1} + {\alpha _2}} \right)}}$, $\delta=2/\xi$, and $\Gamma(\cdot,\cdot)$ is the upper incomplete Gamma function.
\begin{proof}
Based on~\eqref{Power for sensing}, the power outage probability of spectrum sensing can be expressed as
\begin{align}\label{optimal power outage_proof_1}
P_s ^{out}&=\Pr \left\{ {P_{{H_1}} \le {P_s}} \right\}\nonumber\\
 &= {E_{{\Phi _p}}}\left\{ {\prod\limits_{p \in {\Phi _p},\left\| {{d_p}} \right\| \ge {d_0}} {\Pr \left\{ {{{\left\| {{{\bf{h}}_p}} \right\|}^2} \le d_p^\xi {\mu _s}} \right\}} } \right\}\nonumber\\
&={{E_{{\Phi _p}}}\left\{ {\prod\limits_{p \in {\Phi _p},\left\| {{d_p}} \right\| \ge {d_0}} {{F_{{{\left\| {{{\bf{h}}_p}} \right\|}^2}}}\left( {d_p^\xi {\mu _s}} \right)} } \right\}},
\end{align}
where ${F_{{{\left\| {{\mathbf{h}_p}} \right\|}^2}}}$ is the cumulative density function (CDF) of ${{\left\| {{\mathbf{h}_p}} \right\|}^2}$ and is given by
\begin{align}\label{CDF h_p}
{F_{{{\left\| {{\mathbf{h}_p}} \right\|}^2}}}\left( x \right) = 1 - {e^{ - x}}\left( {\sum\limits_{m = 0}^{M - 1} {\frac{{{x^m}}}{{m!}}} } \right).
\end{align}
Applying the moment generating function, we rewrite \eqref{optimal power outage_proof_1} as
\begin{align}\label{optimal power outage_proof_2}
P_s ^{out}=\exp \left[ { - {\lambda _p}\int\limits_{{R^2}} {\left( {1 - {F_{{{\left\| {{{\bf{h}}_p}} \right\|}^2}}}\left( {d_p^\xi {\mu _s}} \right)} \right)} d{d_p}} \right].
\end{align}
Then changing to polar coordinates and substituting \eqref{CDF h_p} into \eqref{optimal power outage_proof_2}, we obtain
\begin{align}\label{optimal power outage_proof_3}
P_s ^{out}=\exp \left[ { - 2\pi {\lambda _p}\sum\limits_{m = 0}^{M - 1} {\frac{{\mu _s^m\int_{{d_0}}^\infty  {{d_p}^{m\xi  + 1}{e^{ - {\mu _s}d_p^\xi }}} d{d_p}}}{{m!}}} } \right].
\end{align}
Applying \cite[ Eq. (3.381.9)]{gradshteyn} to calculate the integral, we obtain \eqref{optimal power outage}.

Similarly, based on \eqref{power for transmission}, the power outage probability of data transmission $P_t ^{out}$ can be expressed as follows:
\begin{align}\label{power outage t}
P_t ^{out}&=\Pr \left\{ {P_{{T_2}} \le {P_t}} \right\}\nonumber\\
&=\Pr \left\{ {\mathop {\max }\limits_{{p \in {\Phi _p},\left\| {{d_p}} \right\| \ge {d_0}}} \left\{ {{{\left\| {{\mathbf{h}_p}} \right\|}^2}{d_p}^{ - \xi }} \right\} \le \mu_t } \right\}.
\end{align}
Following the similar procedure as \eqref{optimal power outage_proof_3} and applying  ${\mu _t} \to {\mu _s}$, we obtain $P_t ^{out}$ in \eqref{optimal power outage}.

The proof is completed.
\end{proof}
\end{theorem}

\subsection{Compressive spectrum sensing}
In the scenario of CS based single SU with WPT, the original signals can be obtained by solving the $l_1$ norm optimization problem as follows:
\begin{align}
&\min {{\left\| {{{s}_j}\left( f \right)} \right\|}_1}, \nonumber\\
&s.t.\left\| {\Theta_j  \cdot {{ s}_j}\left( f \right) - {x_j}} \right\|_2^2 \le {\varepsilon _j},
\label{single_node_recovery}
\end{align}
where ${{\varepsilon _j}}$ is the error bound related to the noise level. This optimization problem can be solved by many existing algorithms for CS, such as by adopting many existing algorithms such as CoSaMP~\cite{needell2009cosamp}, SpaRCS~\cite{waters2011sparcs} etc.

The performance metric of spectrum sensing at single SU can be measured by the probability of detection $P_d$ and probability of false alarm $P_f$. For a target probability of detection, ${{\bar P}_d}$, with which the PUs are sufficiently protected, the threshold can be determined by~\eqref{threshold} accordingly if the number of samples $n$ is fixed. As a result, $P_f$ at a single SU can be derived as follows:
\begin{align}\label{pf}
{P_f} =& Q\left( {{Q^{ - 1}}\left( {{{\bar P}_d}} \right)\sqrt {\frac{{\sigma _s^2 + {\sigma ^2}}}{{{\sigma ^2}}}}  + \sqrt {\frac{n}{2}} \frac{{\sigma _s^2}}{{{\sigma ^2}}}} \right),
\end{align}
where ${\sigma^2}$ and ${\sigma _s^2}$ refer to the noise power and signal power, respectively.

Assuming the energy cost per sample ${e_s}$ in spectrum sensing is fixed, the energy consumption of spectrum sensing is proportional to the number of collected samples as given by:
\begin{align}\label{N}
n = \frac{{\beta T {P_s}}}{{{e_s}}}.
\end{align}
In fact, the number of collected samples $n$ is determined by the sensing time slot. In this case, the energy consumed by reporting collected measurements and receiving decision results between SUs to the FC is ignored. Substituting \eqref{N} into \eqref{pf}, we obtain
\begin{align}\label{pf_2}
{P_f} = \left( {1 - Q\left( {{Q^{ - 1}}\left( {{{\bar P}_d}} \right)\left( {1 + \frac{{\sigma _s^2}}{{{\sigma ^2}}}} \right) + \sqrt {\frac{{\beta T{P_s}}}{{2{e_s}}}} \frac{{\sigma _s^2}}{{{\sigma ^2}}}} \right)} \right).
\end{align}

\subsection{Throughput analysis}
By considering the power outage probability, the throughput at the single SU in the CR networks can be expressed as:
\begin{align}\label{throughput}
\tau =& \left( {1 - P_s^{out}} \right) \times \left( {1 - P_t^{out}} \right) \times \left( {1 - {P_f}} \right) \nonumber \\
 &\times\left( {1 - {\alpha _1} - \beta  - {\alpha _2}} \right) {\tau_t},
\end{align}
where ${\tau_t} = {\log _2}\left( {1 + \frac{{{P_t}}}{{{N_0}}}} \right)$ is the throughput for the data transmission slot, and ${N_0}$ refers to the AWGN level in the data transmission channel. Here we simplify the data transmission process and do not consider the fading, which means the throughput ${\tau_t}$ is only determined by the transmit signal-to-noise-radio (SNR).

When implementing CS technique to achieve sub-Nyquist sampling rate at an SU, It has been proven that the exact signal recovery can be guaranteed if the number of collected measurements satisfies $\Lambda \ge C \cdot K\log \left( {{n \mathord{\left/
 {\vphantom {n K}} \right.
 \kern-\nulldelimiterspace} K}} \right)$, where $C$ is some constant depending on each instance~\cite{cande2008introduction}. If the signal is recovered successfully by $\Lambda$ samples, the achieved $P_f$ would be the same as that no CS technique is implemented with $n$ samples. Therefore, the necessary sensing time slot to achieve the same $P_f$ can be reduced to $\kappa \beta $ with CS implemented. Replacing $P_s^{out}$, $P_t^{out}$ and $P_f$ in~(\ref{throughput}) by~(\ref{optimal power outage}) and~\eqref{pf_2} respectively, the full expression of the throughput with CS implemented can be expressed as follows:
\begin{align}\label{throughput_full}
\tau_{cs} =& \prod\limits_{\zeta  = \left\{ {s,t} \right\}} {\left( {1 - \exp \left( { - \frac{{\pi {\lambda _p}\delta }}{{{{\left( {\mu _\zeta ^{cs}} \right)}^\delta }}}\sum\limits_{m = 0}^{M - 1} {\frac{{\Gamma \left( {m + \delta ,\mu _\zeta ^{cs}d_0^\xi } \right)}}{{m!}}} } \right)} \right)}\nonumber\\
&\times\left( {1 - Q\left( {{Q^{ - 1}}\left( {{{\bar P}_d}} \right)\left( {1 + \frac{{\sigma _s^2}}{{{\sigma ^2}}}} \right) + \sqrt {\frac{{\beta T{P_s}}}{{2{e_s}}}} \frac{{\sigma _s^2}}{{{\sigma ^2}}}} \right)} \right)\nonumber\\
&\times\left( {1 - {\alpha _1} - \kappa \beta  - {\alpha _2}} \right) \times {\log _2}\left( {1 + \frac{{{P_t}}}{{{N_0}}}} \right).
\end{align}

If there is no CS technique implemented at SUs, the time slot fraction for spectrum sensing follows the condition that $0 \le \beta  \le 1$. When the CS technique is implemented at SUs, the time slot for spectrum sensing follows the condition that $C K\log \left( {{n \mathord{\left/
 {\vphantom {n K}} \right.
 \kern-\nulldelimiterspace} K}} \right) \le \frac{{\kappa \beta T{P_s}}}{{{e_s}}} \le n$. By combining these two conditions, the constraint for $\beta$ becomes $\frac{{{e_s}\left( {CK\log \left( {{n \mathord{\left/
 {\vphantom {n K}} \right.
 \kern-\nulldelimiterspace} K}} \right)} \right)}}{{\kappa T{P_s}}} \le \beta  \le 1$. With the implementation of CS, the $\beta$ in ~\eqref{optimal power outage} is replaced by $\kappa \beta$.

Furthermore, the throughput can be maximized by solving the following problem:
\begin{align}\label{throughput_optimization}
\left( {{\rm{P}}0} \right):   &\begin{array}{*{20}{c}}
{ \mathop {\max }\limits_{{\alpha _1},\beta ,{\alpha _2},{P_t}} }&{{\tau _{cs}}},
\end{array}\\ \nonumber
 s.t.  {\kern 1pt}{\kern 1pt}{\kern 1pt}& {C_1}:0 \le {\alpha _1} \le 1,\\ \nonumber
& {C_2}:\frac{{{e_s}\left( {CK\log \left( {{n \mathord{\left/
 {\vphantom {n K}} \right.
 \kern-\nulldelimiterspace} K}} \right)} \right)}}{{\kappa T{P_s}}} \le \beta  \le 1,\\ \nonumber
& {C_3}:{\alpha _{2,\min }} \le {\alpha _2} \le 1,\\ \nonumber
& {C_4}:0 \le 1 - {\alpha _1} - \kappa \beta  - {\alpha _2} \le 1,\\ \nonumber
& {C_5}:{P_{t,\min }} \le {P_t} \le {P_{t,\max }},
\end{align}
where ${\alpha _{2,\min }}$ refers to the minimum time slot fraction for signal recovery at the FC and data transmission between the SU and FC. $P_{t,\min }$ and $P_{t,\max }$ refer to the allowed minimum and maximum power levels for data transmission period. It is noticed that~\eqref{throughput_optimization} is a constrained nonlinear optimization problem and the objective function is very complex. However, the constraints are linear, which motives us to solve the optimization problem by the following three methods:

\begin{enumerate}
  \item \emph{Grid search}: The grid search algorithm for solving~\eqref{throughput_optimization} can be described in Algorithm~\ref{grid_search}. The grid search algorithm can find out the global optimal value if the step size ${\Delta _i}\left( {i = 1,2,3,4} \right)$ for ${\alpha _1}$, $\beta $, ${\alpha _2}$, ${P_t}$ are small enough. However, the computational complexity would greatly increase if the step is set to be small enough.
  \item \emph{fmincon}: \emph{fmincon} is a toolbox in MATLAB which is efficient but it may return a local optimal value.
  \item \emph{Random sampling}: A set $S = \left\{ {{v_1},{v_2}, \cdots ,{v_{\rm Z}}} \right\} $ that satisfies the conditions in~\eqref{throughput_optimization} is generated randomly, where ${v_i} = \left( {{\alpha _1},\beta ,{\alpha _2},{P_t}} \right)$ ($i \in \left\{ {1,2, \cdots ,{\rm Z}} \right\}$) is a tuple of generated random samples, and ${\rm Z}$ is the number of generated tuples. The accuracy of this method depends on number of tuples ${\rm Z}$ generated for calculation. This method is efficient for solving \eqref{throughput_optimization} as it does not rely on advanced optimization techniques and the method of generating $\left( {{\alpha _1},\beta ,{\alpha _2},{P_t}} \right)$ is efficient.
\end{enumerate}

\begin{algorithm}[!t]
\caption{Grid search}
\label{grid_search}
\begin{algorithmic}[1]
\STATE Initialization: ${\Delta _1}$, ${\Delta _2}$, ${\Delta _3}$, ${\Delta _4}$, and ${\tau _{temp}} = \emptyset $.

\FORALL  {${\alpha _1} \in \left( {0:{\Delta _1}:1} \right)$, $\beta  \in \left( {\frac{{{e_s}\left( {CK\log \left( {{n \mathord{\left/
 {\vphantom {n K}} \right.
 \kern-\nulldelimiterspace} K}} \right)} \right)}}{{\kappa T{P_s}}}:{\Delta _2}:1} \right)$, ${\alpha _2} \in \left( {{\alpha _{2,\min }}:{\Delta _3}:1} \right)$, ${P_t} \in \left( {{P_{t,\min }}:{\Delta _4}:{P_{t,\max }}} \right)$}
\WHILE {${0 \le 1 - {\alpha _1} - \kappa \beta  - {\alpha _2} \le 1}$}
\STATE ${\tau _{temp}} = [{\tau _{temp}},{\tau _{cs}}]$ where ${\tau _{cs}}$ is expressed as~\eqref{throughput_full}.
\ENDWHILE
\ENDFOR
\STATE Return $\tau_{max}={\tau _{temp}}$.
\end{algorithmic}
\end{algorithm}

\section{Throughput Analysis of Multiple Secondary Users}
In this section, the closed-form expressions for the power outage probability of data transmission are deprived for both the active and inactive SUs, respectively. After considering the throughput optimization of a single SU locally in Section III, we optimize the throughput of the CSS networks with multiple SUs including the active and inactive ones with implementing MC technique.

\subsection{Power outage probability analysis}
In this subsection, we provide power outage probability analysis for spectrum sensing and data transmission for both active and inactive SUs. In the considered CSS networks, the number of active participating SUs is $J_1$ and the number of inactive SUs is $J_2$.
\subsubsection{Power Outage Probability of Spectrum Sensing} Note that the power outage probability of sensing is always zero. This behavior can be explained as follows: 1) For active SUs, it is assumed that the harvested energy from the first time slot is enough for them to take measurements for spectrum sensing. Therefore, the power outage probability of spectrum sensing for active SUs are zero; and 2) For the inactive SUs, they will not perform spectrum sensing and only transmit data. As a consequence, there is no spectrum sensing outage.
\subsubsection{Power Outage Probability of Data Transmission} For those active SUs, the power outage probability of data transmission is same that of the single SU scenario. For those inactive SUs, as all the time slots before data transmission are used for energy harvesting, the power outage probability is different as that of active SUs. The following theorem provides the exact analysis for the power outage probability of data transmission for both active and inactive SUs in CSS networks.
\begin{theorem}
The power outage probability of data transmission at the active and inactive SUs in the fourth time slot can be expressed in closed-form as follows:
\begin{align}\label{optimal power outage css}
P_\psi ^{out} = \exp \left[ { - \frac{{\pi {\lambda _p}\delta }}{{\mu _\psi ^\delta }}\sum\limits_{m = 0}^{M - 1} {\frac{{\Gamma \left( {m + \delta ,{\mu _\psi }d_0^\xi } \right)}}{{m!}}} } \right]
\end{align}
where $\psi  \in \left( {a,i} \right)$, ${\mu _a} = \frac{{{P_t}\left( {1 - {\alpha _1} - \beta  - {\alpha _2}} \right) + {P_s}\beta }}{{\eta {P_p}A\left( {{\alpha _1} + {\alpha _2}} \right)}}$, and ${\mu _i}~=~\frac{{{P_t}\left( {1 - {\alpha _1} - \beta  - {\alpha _2}} \right)}}{{\eta {P_p}A\left( {{\alpha _1} + \beta  + {\alpha _2}} \right)}}$.
\begin{proof}
Based on~\eqref{PH3}, the power outage probability of data transmission at the active SUs is the same as $P_t^{out}$ as given in \eqref{optimal power outage} by replacing ${\mu _a} \to {\mu _\zeta }$.

Similarly, based on~\eqref{PH3}, the power outage probability of data transmission at the inactive SUs can be expressed as follows:
\begin{align}\label{optimal power outage_proof_1_css}
P_i ^{out}& = \Pr \left\{ {{P_{{H_3}}} < {P_t}} \right\}\\ \nonumber
& = \Pr \left\{ {\mathop {\max }\limits_{p \in {\Phi _p},\left\| {{d_p}} \right\| \ge {d_0}} \left\{ {{{\left\| {{{\bf{h}}_p}} \right\|}^2}{d_p}^{ - \xi }} \right\} < {\mu _i}} \right\}.
\end{align}
Following the similar procedure as \eqref{optimal power outage_proof_3} and applying ${\mu _i} \to {\mu _s}$, we can obtain $P_{i} ^{out}$ in \eqref{optimal power outage css}.

The proof is completed.
\end{proof}
\end{theorem}

\subsection{Matrix completion based cooperative spectrum sensing}
As the spectrum is normally underutilized in practice, the transmitted signals exhibit a sparse property in frequency domain, which can be transformed into a low-rank property of the matrix $X$ at the FC. When only the $J_1$ active SUs send compressed samples to the FC, the matrix with collected samples is incomplete at the FC. The exact matrix can be obtained by solving the following low-rank MC problem:
\begin{align}
\begin{array}{l}
\min {\kern 1pt} {\kern 1pt}{\kern 1pt} {\kern 1pt}{\kern 1pt} {\kern 1pt} {\rm{rank}}\left( {S\left( f \right)} \right)\\
\rm{s.t.}{\kern 1pt} {\kern 1pt}{\kern 1pt}{\kern 1pt} {\kern 1pt}{\kern 1pt} {\kern 1pt}\left\| {{\rm{vec}}\left( {\Theta  \cdot \emph{S}\left( \emph{f} \right)} \right) - {\rm{vec}}\left( X \right)} \right\|_2^2 \le {\boldsymbol{\varepsilon}}
\end{array},
\label{rank_min}
\end{align}
where ${\rm{rank}}\left(  \cdot  \right)$ is the rank function of a matrix whose value is equal to the number of nonzero singular values of the matrix, and ${\boldsymbol{\varepsilon}}  = \left[ {{\varepsilon _1}, \ldots ,{\varepsilon _J}} \right]$. However, the problem in~(\ref{rank_min}) is NP-hard due to the combinational nature of the function ${\rm{rank}}\left(  \cdot  \right)$. It has been proved that the nuclear norm is the best convex approximation of the rank function over the unit ball of matrixes with norm no less than one~\cite{Fazel_matrix:2002}. Therefore, (\ref{rank_min}) can be replaced by the following convex formulation:
\begin{align}\label{low-rank}
\begin{array}{l}
\min {\kern 1pt} {\kern 1pt}{\kern 1pt} {\kern 1pt} {\left\| {S\left( f \right)} \right\|_ * }\\
\rm{s.t.}{\kern 1pt} {\kern 1pt}{\kern 1pt} {\kern 1pt}{\kern 1pt} {\kern 1pt}{\kern 1pt} \left\| {{\rm{vec}}\left( {\Theta  \cdot \emph{S}\left( \emph{f} \right)} \right) - {\rm{vec}}\left( X \right)} \right\|_2^2 \le {\boldsymbol{\varepsilon}}
\end{array}.
\end{align}
The problem in~\eqref{low-rank} such as singular value decomposition~\cite{drineas2002competitive}, nuclear norm minimization~\cite{candes2009exact} and so on.

Once the exact matrix is recovered, energy detection can be used to determine the spectrum occupancy. In the CSS networks, as all the participating SUs send the collected samples to the FC, the data fusion is considered. Suppose the channel coefficients from the PUs to each participating SUs are known. When the channel coefficients are unknown, the weighting factor associated with the $j$th SU is set to be ${g_j} = \frac{1}{{\sqrt J }}$. By using the maximal ratio combining, probability of false alarm of the CSS networks $Q_f$ becomes
\begin{align}\label{Qf}
{Q_f}=Q\left( {{Q^{ - 1}}\left( {{{\bar P}_d}} \right)\sqrt {1 + \frac{{\sigma _s^2}}{{{J}{\sigma ^2}}}\rm H }  + \sqrt {\frac{n}{{2{J}}}} \frac{{\sigma _s^2}}{{{\sigma ^2}}}\rm H} \right).
\end{align}
where ${\rm H} = \sum\limits_{j = 1}^J {{{\left| {{h_j}} \right|}^2}} $ refers to the channel coefficients for all the participating SUs in the considered CSS networks. In order to simplify the case, we choose $h_j=1$. Consequently, $Q_f$ can be expressed as
\begin{align}\label{Qf_short}
{Q_f} = Q\left( {{Q^{ - 1}}\left( {{{\bar P}_d}} \right)\sqrt {1 + \frac{{\sigma _s^2}}{{{\sigma ^2}}}}  + \sqrt {\frac{{nJ}}{2}} \frac{{\sigma _s^2}}{{{\sigma ^2}}}} \right).
\end{align}

\subsection{Throughput analysis}
By applying the MC technique at the FC, $\beta $ in~\eqref{optimal power outage css} should be replaced by $\kappa \beta$. In addition, the throughput of considered CSS networks with multiple SUs can be expressed as
\begin{align}\label{throughput_css}
{\tau _{mc}} =& \sum\limits_{j = 1}^J {\left( {1 - P_\psi ^{out}\left( j \right)} \right) \times \left( {1 - {Q_f}} \right)}\nonumber\\
&\times \left( {1 - {\alpha _1} - \kappa \beta  - {\alpha _2}} \right){\tau _t},
\end{align}
where $\kappa$ is defined as the compression ratio at active SUs. $\psi =a$ refers to the $J_1$ number of active SUs and $\psi =i$ refers to the rest of inactive SUs in the considered networks. If no MC technique is implemented at the FC and each SU is supposed to take samples at Nyquist rate, all the participating SUs will send samples to the FC. Then the throughput of the considered networks can be given by replacing $P_\psi ^{out}\left( j \right)$ in~\eqref{throughput_css} with $P_a ^{out}\left( j \right)$ and $\kappa=1$. The constraints for the multiple nodes case are the same as that for the single node case except the condition for $\beta$. In the multiple nodes case, the condition for $\beta$ follows $C{\mu ^2}\nu K{\log ^6}\nu  \le \frac{{\kappa \beta T{P_s}{J_1}}}{{{e_s}}} \le nJ$. Combining the condition for $\beta$ in case of no MC implemented, $\beta$ should satisfy $\frac{{{e_s}\left( {C{\mu ^2}\nu K{{\log }^6}\nu } \right)}}{{\kappa T{P_s}{J_1}}} \le \beta  \le 1$.

The throughput can be maximized by solving the following problem:
\begin{align}\label{throughput_optimization_css}
\left( {{\rm{P}}1} \right):   &\begin{array}{*{20}{c}}
{ \mathop {\max }\limits_{{\alpha _1},\beta ,{\alpha _2},{P_t}} }&{{\tau _{mc}}}
\end{array}\\ \nonumber
{{\rm{s}}.{\rm{t}}.}{\kern 1pt}{\kern 1pt}{\kern 1pt}{\kern 1pt}{\kern 1pt}{\kern 1pt} &{{C_1},{C_3},{C_4}{\kern 1pt}{\rm{and}}{\kern 1pt}{\kern 1pt}{C_5},}\\ \nonumber
&{C_6}:\frac{{{e_s}\left( {C{\mu ^2}\nu k{{\log }^6}\nu } \right)}}{{\rho T{P_s}{J_1}}} \le \beta  \le 1,
\end{align}
where $C{\mu ^2}\nu k{\log ^6}\nu $ is the lower bound of the number of observed measurements at the FC to guarantee the exact matrix recovery~\cite{candes2010matrix}. Here, $\nu = \max \left( {n,J} \right)$ and $\mu  = {\rm O}\left( {\sqrt {\log \nu } } \right)$. It is noticed that the structure of $\left( {{\rm{P}}1} \right)$ is similar as $\left( {{\rm{P}}0} \right)$. Thus we can use the similar methods to obtain the optimal throughput for the CSS network.

\section{Numerical Results}\label{Numerical Results}
In the simulation, we set the frame period to be $T=1$ s, the transmit power of PBs to be $P_p=43$ dBm, the number of antennas of PB to be $M=32$, the carrier frequency for power transfer to be $900$ MHz, and the energy conversion efficiency of WPT to be $\eta=0.8$. In addition, it is assumed that the target probability of detection ${{\bar P}_d}$ is $90\%$. In this paper, we let $d_0\geq 1$~m to make sure the path loss of WPT is equal or greater than one.  In the following part, $\rm{SNR} = \frac{{\sigma _s^2}}{{{\sigma ^2}}}$ refers to the SNR in sensing channels.

\subsection{Numerical results on optimizing throughput of single secondary user}
In this subsection, simulation results of the optimized throughput of single SU are demonstrated after the derived power outage probability in~(\ref{optimal power outage}) and $P_f$ in~(\ref{pf_2}) are verified by Monte Carlo simulations.

\begin{figure}[t!]
    \begin{center}
        \includegraphics[width=3.8in,height=2.5in]{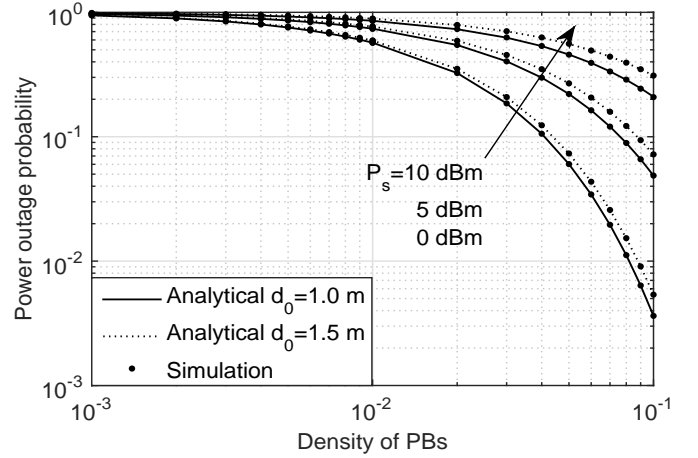}
        \caption{Power outage probability of spectrum sensing versus density of PBs with $M=32$, $P_p=43$ dBm, $\alpha_1=0.25$, $\alpha_2=0.2$, and $\beta=0.25$.}
        \label{power outage density}
    \end{center}
\end{figure}

Fig.~\ref{power outage density} plots the power outage probability of spectrum sensing versus density of PBs with different power threshold $P_{s}$. The black solid and dash curves are used to represent the analytical results with $d_0=1$ m and $d_0=1.5$ m, respectively, which are both obtained from~\eqref{optimal power outage}. Monte Carlo simulations are marked as ``$\bullet$" to verify our derivation. The figure shows the precise agreement between the simulation and analytical curves. One can be observed is that as density of PBs increases, the power outage probability dramatically decreases. This is because the multiuser diversity gain is improved with increasing number of PBs when charging with WPT. The figure also demonstrates that the outage occurs more frequently as the power threshold $P_{s}$ and the radius of protection zone $d_0$ increase.

\begin{figure}[t!]
    \begin{center}
        \includegraphics[width=3.8in,height=2.5in]{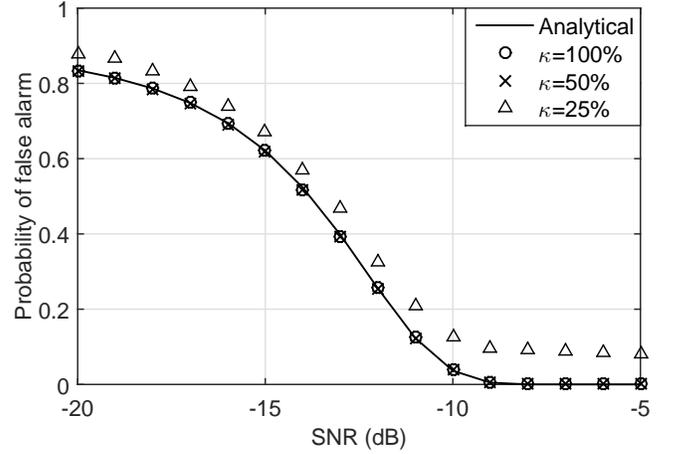}
        \caption{Probability of false alarm versus SNR in sensing channels with different compression ratio $\kappa$, ${\bar P_d}=90\%$, and sparsity level $= 12.5\%$.}
        \label{false alarm probability}
    \end{center}
\end{figure}

Fig.~\ref{false alarm probability} plots the probability of false alarm  $P_f$ versus SNR in sensing channels with different compression ratio $\kappa$. It is observed that $P_f$ decreases with higher SNR, which would improve the throughput of secondary network. The black solid curve is used to represent the analytical result which is obtained from ~\eqref{pf_2} with enough protection to PUs being provided. Monte Carlo simulations with compression ratio $\kappa=100\%$ are marked as ``$\circ$" to verify our derivation, which represents the scenario without CS technique implemented. The figure shows precise agreement between the simulation and analytical curves. When $\kappa$ is reduced to 50\%, it is noticed that $P_f$ is still well matched with the analytical result, which means the performance of spectrum sensing would not be degraded when only 50\% of the samples are collected at an SU. When $\kappa$ is further reduced to 25\%, $P_f$ is increased, which means the signal recovery is not exact any more. As a result, throughput of the secondary network would be degraded correspondingly. Actually, the lowest compression ratio for successful signal recovery is dependent on the sparsity levels of received signals and the available prior information. This is out of the scope of this paper and would not be discussed further.

\begin{figure}[t!]
    \begin{center}
        \includegraphics[width=3.8in,height=2.5in]{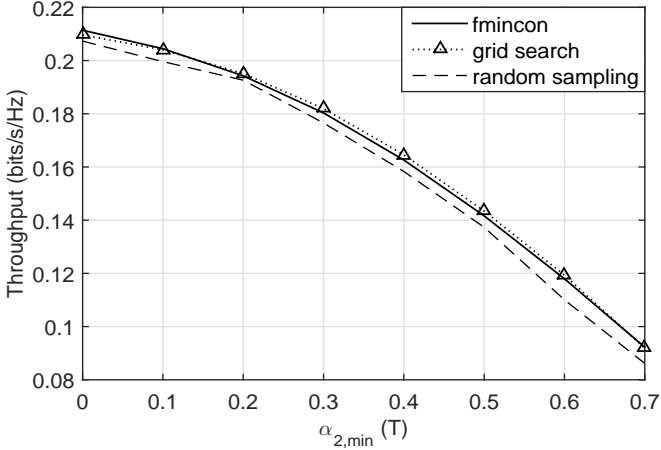}
        \caption{Throughput of single SU $\tau_{cs}$ versus lower bound of the third time slot $\alpha _{2,\min }$, $SNR=-10$ dB, and $\kappa=1$.}
        \label{throughput_optimized_random}
    \end{center}
\end{figure}

Fig.~\ref{throughput_optimized_random} plots the achievable throughput $\tau_{cs}$ versus the lower bound of time allocated to the third slot $\alpha_2$. Here $\alpha_2$ is reserved for signal recovery at the FC and data transmission between the SU and the FC. In this figure, several observations are drawn as follows: 1) The maximal throughput achieved by grid search method is slight higher than that of \emph{fmincon} method as \emph{fmincon} relies on the initial input and may return a local optimal value. However, the accuracy of grid search method is dependent on the step sizes; 2) The random sampling method achieves lower throughput than grid search and \emph{fmincon }methods, which demonstrates the benefits of the presented grid search and fmincon methods. When the generated sets increase for random sampling, the achieved maximal throughput get closer to the optimal but the computational complexity is much increase; 3) It is seen that the optimal value of time assigned to the third slot $\alpha _{2}$ always equals to the lower bound $\alpha _{2,\min }$. This gives a sign that the throughput can be improved if the time slot for the signal recovery at the FC and data transmission between SUs and the FC is reduced. In other words, the energy harvesting should be done mainly in the first time slot $\alpha_1$ to reduce the power outage probability in the following spectrum sensing slot if the signal recovery and data transmission between the SU and FC can be promised.

\begin{figure}[t!]
    \begin{center}
        \includegraphics[width=3.8in,height=2.5in]{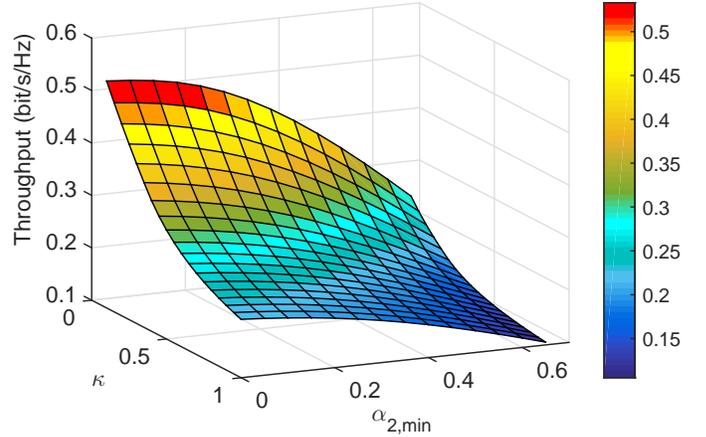}
        \caption{Optimized throughput of single SU $\tau_{cs}$ versus lower bound of the third time slot $\alpha _{2,\min }$ and compression ratio $\kappa$, and $SNR=-10$ dB in sensing channels.}
        \label{throughput_optimized_3d}
    \end{center}
\end{figure}

Fig.~\ref{throughput_optimized_3d} plots the achieved optimal throughput $\tau_{cs}$ versus lower bound $\alpha _{2 }$ and compression ratio $\kappa$ when solving the problem in~\eqref{throughput_optimization}. It shows that the achieved maximal throughput increases with decreasing $\kappa$ and increasing $\alpha _{2,\min }$. This behavior can be explained as follows: as $\kappa$ decreases, the number of samples to be collected for spectrum sensing at an SU is reduced as the signal in original size of $N \times 1$ can be recovered from less number of $\Lambda $ measurements by utilizing CS technique. When the time slot assigned for spectrum sensing is reduced, the energy consumption for spectrum sensing is reduced. As a result, the time which can be assigned for data transmission is increased. Therefore, the throughput of the secondary network is improved. By optimizing the transmission power $P_t$, the energy harvested in the current frame period would be fully utilized and the maximal throughput can be achieved accordingly. It should be noted that when the compression ratio $\kappa$ is set to 100\% and the lower bound of the third time slot $\alpha_{2,min}$ is zero, the achieved throughput can be regarded as that of the traditional frame structure design without considering sub-Nyquist sampling. The case when $\kappa$ is set to 100\% can be regarded as a benchmark for the performance metric.

\subsection{Numerical results on optimizing throughput of multiple secondary users}
In the case of optimizing the throughput of the CSS networks, the total number of participating SUs is set be to $J=50$, including both the active and inactive SUs. Comparing the format of (P1) and (P0), we can see that both of them are linear constrained. Therefore, similar as (P0), the grid search method can be applied to obtain the optimal throughput but with non-negligible complexity, especially for the case of optimizing throughput of the whole cooperative network. The \emph{fmincon} method can be adopted to obtain the sub-optimal throughput efficiently. In the following simulations, the \emph{fmincon} method is utilized to solve the optimization problem (P1). In addition, as aforementioned in the introduction part, many algorithms have been proposed for the low-rank MC based cooperative spectrum sensing. With $\hat P_d=90\%$, the detection performance with different compression ratios is presented in the authors' previous work in~\cite{Zhijin_TSP:2015}, which would not be demonstrated here again to reduce redundancy. In the following simulations, how the achieved throughput is influenced by parameters, such as the number of active SUs $J_1$, compression ratio $\kappa$ and the lower bound of the third time slot $\alpha _{2,\min }$, would be demonstrated.

\begin{figure}[t!]
    \begin{center}
        \includegraphics[width=3.8in,height=2.5in]{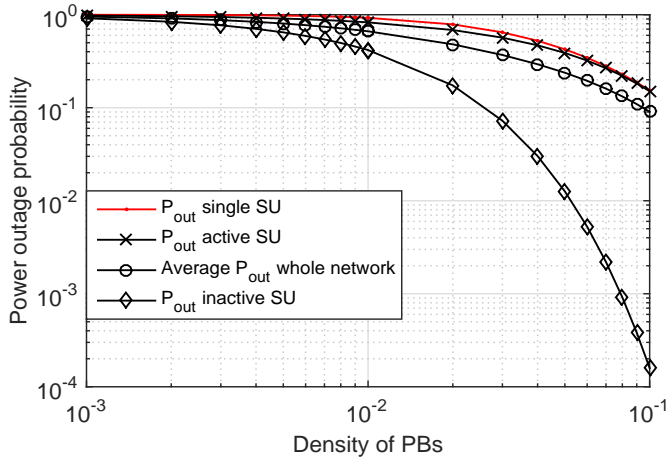}
        \caption{Power outage probability comparison for single SU and multiple SUs with versus density of PBs, $P_s=0$ dBm, $\alpha_1=0.25$, $\alpha_2=0.20$, and $\beta=0.25$.}
        \label{Pout_single_multiple}
    \end{center}
\end{figure}

Fig.~\ref{Pout_single_multiple} plots the power outage probability ${P_{out}}$ versus density of PBs with different power threshold $P_{s}$. In this case, the power outage probability here is for the whole system, which can be calculated as ${P_{out}} = 1 - \left( {1 - P_{out}^s} \right) \times \left( {1 - P_{out}^t} \right)$. Both the single SU scenario and multiple SUs scenario are illustrated in the figure. It can be observed that as density of PBs increases, the power outage probability dramatically decreases, which is caused by that the multiuser diversity gain is improved with increasing number of PBs when charging with WPT. We can also observe that the ${P_{out}}$ of  multiple SUs scenario is lower than that of single SU scenario. This is because the power outage probability of spectrum sensing is always zero in multiple SUs scenario, which in turn lower the power outage probability of the whole system. For the multiple SUs scenario, the ${P_{out}}$ of the active SUs, inactive SUs and the average ${P_{out}}$ of the CSS networks are all presented in the figure. It is noted the averaged ${P_{out}}$ falls between the ${P_{out}}$ of active SUs and inactive SUs, which as we expected.

\begin{figure}[t!]
    \begin{center}
        \includegraphics[width=3.8in,height=2.5in]{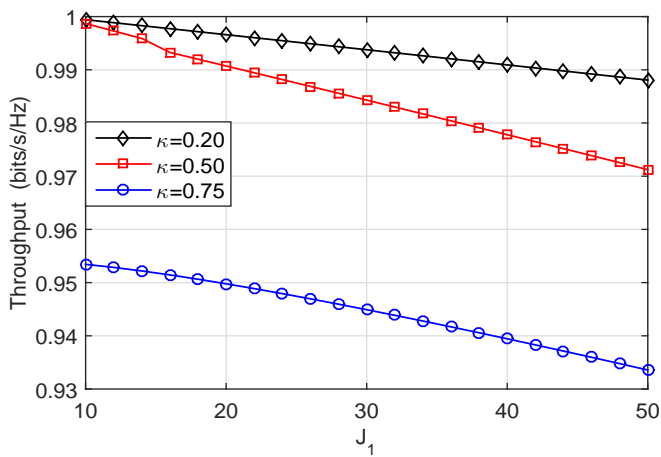}
        \caption{Optimized throughput averaged on per SU of multiple SUs $\tau_{mc}$ versus number of active SUs $J_1$ and compression ratio $\kappa$, $SNR=-10$~dB in sensing channels, and $\alpha _{2,\min }=0.05$.}
        \label{averaged_throughput_different_kappa_J1}
    \end{center}
\end{figure}

Fig.~\ref{averaged_throughput_different_kappa_J1} plots the throughput averaged on per SU $\tau_{mc}$ with different number of active SUs $J_1$ and different compression ratios $\kappa$. In this case, it is assumed that the exact matrix completion can be guaranteed when the number of active SUs $J_1$ is in the range of 10 to 50 with compression ratio $\kappa$ changes from 20\% to 100\%. It shows that the average throughput achieves the best performance when the number of active SUs $J_1$ is set to be the minimal number in comparison with that of case $J_1=J$. This benefits from the non-active SUs, which can save energy for spectrum sensing and harvest more energy for data transmission. With compression ratio $\kappa$ decreasing from 75\% to 20\%, the achieved optimal throughput is increased as the necessary time slot for spectrum sensing is reduced.

\begin{figure}[t!]
    \begin{center}
        \includegraphics[width=3.8in,height=2.5in]{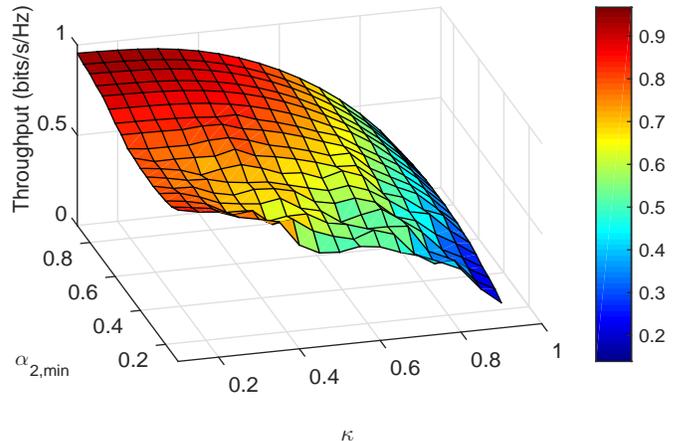}
        \caption{Optimized throughput averaged on per SU of multiple SUs $\tau_{mc}$ versus lower bound $\alpha _{2,\min }$ and compression ratio $\kappa$, and $SNR=-10$~dB in sensing channels.}
        \label{3D_fmincon_different_alpha_2_min_kappa}
    \end{center}
\end{figure}

Fig.~\ref{3D_fmincon_different_alpha_2_min_kappa} plots the achieved maximum throughput averaged on per SU versus different compression ratio $\kappa$ and lower bound for the third time slot $\alpha_{2,min}$. In this case, the number of active SUs is $J_1=30$. The achieved throughput shown in Fig.~\ref{3D_fmincon_different_alpha_2_min_kappa} is based on the condition that the exact matrix recovery can be guaranteed with the given compression ratio. As shown in the figure, the maximum achieved throughput increases with decreasing $\kappa$ and $\alpha_{2,\min}$. The minimal compression ratio guaranteing the exact matrix recovery is nondeterministic, which is dependent on the rank order of the matrix to be recovered. If the rank order $J_1$ is fixed, the larger network size $J$, the lower minimal compression ratio which can guarantee the exact matrix recovery.

\section{Conclusions}
In this paper, a wireless powered cognitive radio network has been considered. In the considered networks, while protecting the primary users, we proposed a new frame structure including energy harvesting, spectrum sensing, energy harvesting and data transmission. In the proposed frame structure, closed-form expressions in terms of power outage probability was derived for the proposed wireless power transfer (WPT) scheme. In addition, sub-Nyquist sampling was performed at SUs to reduce the energy consumption during spectrum sensing. The compressive sensing and matrix completion techniques were adopted at a remote fusion center to perform the signal recovery for detection making on spectrum occupancy. By optimizing the four time slots, throughput of a single secondary user and the whole cooperative networks were maximized, respectively. Simulation results showed that the throughput can be improved in the proposed new frame structure design. We conclude that by carefully tuning the parameters for different time slots and transmit power, WPT can be used along with compressive sensing and matrix completion to provide a high quality of throughput performance for cognitive radio network, with significantly energy computation reduction at power-limited SUs.

\bibliographystyle{IEEEtran}
\bibliography{mybib}

\end{document}